# Quantum Interaction Between Free Electrons and Light Involving First-order and Second-order Process

Hongteng Lin, Xiaotong Xiong, Junjie Liu, Yidong Huang and Fang Liu*

*Department of Electronic Engineering, Tsinghua University, Beijing, China*
*Email: liu_fang@tsinghua.edu.cn*

**Abstract**

Photon-induced Near-field Electron Microscopy (PINEM) effect has revealed the quantum interaction between free electrons and optical near filed, which demonstrated plenty of novel phenomena of manipulating free electron wave packet and detecting/shaping quantum photonic states. However, free electrons generally only absorb/emit one photon at a time, while the physical mechanism and phenomena of free electron-two-photon interaction have not been studied yet. Moreover, the relationship between PINEM and Kapitza-Dirac (KD) effect and nonlinear Compton scattering is still unclear. Here we develop the full quantum theory of electron-photon interaction considering the two-photon process. It is revealed that the emission/absorption of two photons by electrons can be greatly enhanced by manipulating the electric field component of optical near field, and the quantum interference between single-photon and two-photon processes can occur in some circumstances, which affects the photon number state, electron energy states and electron-photon entanglement. Meanwhile, it is found that the KD effect (elastic electron-photon scattering) and nonlinear Compton scattering (inelastic electron-photon scattering) are also a kind of two-photon process and the distribution of electrons can be deduced analytically based on the full quantum theory. Our work uncovers the possible abundant phenomena when free electron interacting with two photons, paves the way for more in-depth studies of nonlinear processes in electron-photon quantum interactions in the future.

## I. Introduction

The interaction between free electrons and light is crucial for exploring the basic physical properties of free electrons and brings fruitful applications. Recently, Photon-induced Near-field Electron Microscopy (PINEM) [1-3] is studied considering the interaction between free electrons and light with quantum-mechanical formalism [4-8] and thus abundant novel phenomena were discovered, such as the optical near-field imaging combining with fs temporal and nm spatial resolution [9-13], the shaping of free electron wavefunction [14-17], the modulation of free electron emission properties and the generation of various quantum light [18-22], etc.

The existing quantum PINEM (Q-PINEM) theory [6-8] only considers the single-photon process, namely the energy exchange between free electron and light is only single photon at a time. Such theoretical model is restricted to analyze single-photon process and cannot be applied for the following physical process and potential applications. (i) Emitting only single photon at a time [23-24] restricts the application for generating various quantum light and entangled photon pairs by free electrons. (ii) Only the longitudinal component of optical near field other than the full vector field can be derived by the electron energy loss spectroscopy (EELS) [6-13]. (iii) What's more, free electrons are also unable to detect directly the phase of the light field [4-5, 8]. (iv) For the manipulation of free electron wave packets using light field, the freedoms of the light field are unable to be fully utilized [25-27]. How can we make a breakthrough on the above limitations of PINEM in theory and establish the analytical model of two-photon process in PINEM?

Besides PINEM effect, the Kapitza-Dirac (KD) effect [28-30] and nonlinear Compton scattering [31-32] are also the phenomena of the interaction between free electrons and light wave, which are considered as the elastic and inelastic scattering of electrons by the ponderomotive potential of light field (namely, electrons are scattered by photon-pairs), respectively. Although the existing theory can be applied for analyzing electron scattering properties [30, 33], there still lacks the full quantum electrodynamics (QED) treatment for KD effect and analytical study for nonlinear Compton scattering to obtain the quantum features of both photons and electrons after interaction. Further, it is natural to raise the question what the intrinsic connection is between PINEM, KD effect and nonlinear Compton scattering.

In this work, we investigate free electron-two-photon interaction systematically by developing the corresponding full quantum theory using scattering matrix method. The universal expression of scattering operator is derived rigorously starting from the interaction Hamiltonian without neglecting the second-order term. It is found that single-photon and two-photon emission/absorption, KD effect and nonlinear Compton scattering can be unified into the derived scattering operator. According to our analytical theory, the electron-two-photon interaction can be enhanced by eight orders of magnitude when using quasi-BIC metasurface. Subsequently, the entanglement features between electron and photons were analyzed and the quantum interference between single-photon process and two-photon process is discovered. Finally, with the derived scattering operator, the KD effect and nonlinear Compton scattering are well explained by deducing the analytical expression of electron transverse momentum probability and electron energy probability distribution, respectively. This work expands the Q-PINEM theory for considering the nonlinear interactions between free electrons and light, and provides a pathway to extend the potential applications of PINEM, e.g., regulating electron-photons entanglement, detecting the full vectors of light field and generating desired quantum light by free electrons.

## II. The theoretical description

**The scattering operator of free electron-light interaction considering two-photon process**

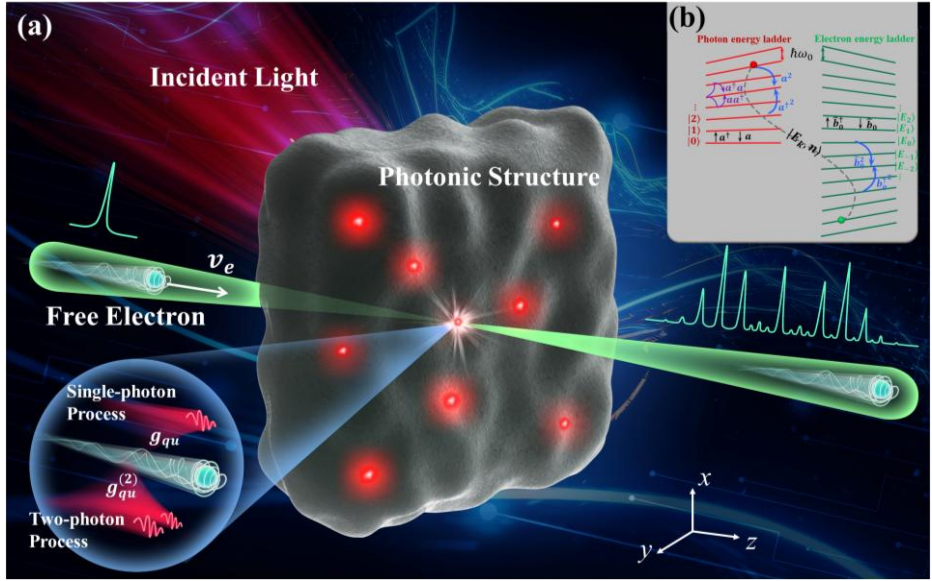

**Fig. 1. The general schematic of free-electron-light interaction when considering two-photon process. (a)** A narrow energy electron beam flies through a certain photonic structure along $z$-axis, quantum interaction occurs between free electron and light, the parameters $g_{qu}$ and $g_{qu}^{(2)}$ are the first-order and second-order quantum coupling constant, respectively, which measures the strength of single-photon and two-photon process in the interaction between free electron and light. **(b)** The inset showing the electron energy ladder and photon energy ladder in electron-light interaction. The electron state is described by infinite energy ladder spacing by light quantum energy $\hbar\omega_0$, while the photonic state is described by a half infinite energy ladder with the same step to electron energy ladder. There depict three situations for electron-light interaction when considering two-photon process, including electron emits/absorbs one photon, electron emits/absorbs two photons at a time and the electron emits (absorbs) a photon and absorbs (emits) a photon simultaneously.

As shown in Fig. 1(a), a focused electron beam, flying over a certain photonic structure along $z$-axis, emits or absorbs a single photon and/or two photons simultaneously when interacting with the optical near field modes or free space modes. The two photons can have the same frequency or different frequencies, while for succinctly and clearly, Fig. 1 depicts the case of the same frequency. The photonic structure could be optical waveguide, cavity, metasurface or even free space, and by designing the optical near field or free space modes, the emission and absorption of two-photon by free electrons could be manipulated.

Similar to quantum PINEM theory considering only single-photon process [2-3, 21], the following approximations are also adopted in this work: (1) The incident monochromatic free electron beam is paraxial with its wave vectors $\bm{k}$ tightly packed around a central wave vector $\bm{k_0}$; (2) After interacting with the optical field effectively, the electrons' wave vector $\bm{k}$ satisfy $|\bm{k} - \bm{k_0}| \ll k_0$, which is also known as nonrecoil approximation; (3) The region that electrons flying through is charge-free, so that the Coulomb gauge, $\varphi(\bm{r}) = 0$, can be taken. Considering the free electron interact with one of the modes in the photonic structure with frequency $\omega_0$, the quantum state can be described as infinite energy ladder for free electron and half infinite energy ladder for photons, respectively, as shown in Fig. 1(b). The energy spacing of both energy ladders is $\hbar\omega_0$, where $\hbar$ is the reduced Planck's constant. Therefore, the photon number state in the photonic structure can be written as $|n\rangle$, where $n$ is the Fock-state index, and the electron state is represented by $|E_k\rangle$ with $E_k = E_0 + k\hbar\omega_0$, where $E_0$ is the initial energy of incident free electron. For $k > 0$, the electron absorbs $k$ photons and gains energy, and for $k < 0$, the electron emits $k$ photons and loses energy. Hence, the joint quantum state of the electron-photon system after interaction is represented by

$$|\psi_f\rangle = \sum_{n=0}^{\infty} \sum_{k=-\infty}^{\infty} c_{nk} |E_k, n\rangle \qquad (1)$$

where $c_{nk}$ is the corresponding coefficient when the photon number state is $|n\rangle$ and the electron energy state is $|E_k\rangle$.

To consider a more comprehensive situation, here we assume that free electron interacts with two different photonic modes simultaneously, whose frequencies are $\omega_1$ and $\omega_2$. Considering the two-photon process, the interaction Hamiltonian under minimal coupling can be written as

$$\widehat{H}_I(\bm{r}) = \sum_{i,j=1,2} \left\{ e\bm{v_e} \cdot \widehat{\bm{A}}_i(\bm{r}) + \frac{e^2}{2m_e\gamma} \left[ \hat{A}_{ix}(\bm{r})\hat{A}_{jx}(\bm{r}) + \hat{A}_{iy}(\bm{r})\hat{A}_{jy}(\bm{r}) + \frac{1}{\gamma^2} \hat{A}_{iz}(\bm{r})\hat{A}_{jz}(\bm{r}) \right] \right\}$$

(2)

where $\boldsymbol{v_e} = v_e \boldsymbol{e_z}$ ($\boldsymbol{e_z}$ is the unit vector in $z$-direction), $m_e$ and $e$ are the velocity of free electron, the rest mass of electron and the electron's elementary charge, respectively, and the Lorentz factor is $\gamma = 1/\sqrt{1 - v_e^2/c^2}$. Here the electromagnetic vector potential $\boldsymbol{A}(\boldsymbol{r})$ and scalar potential $\varphi(\boldsymbol{r})$ are introduced to describe the light field. In previous quantum PINEM theory, the higher order terms of Hamiltonian (e.g. the $A^2$ terms in Eq. (2)) were ignored for considering that free electron only absorbs/emits one photon at the same time [6-7] Although the ponderomotive term was considered for achieving optical-cavity mode squeezing by free electrons [34], the two-photon process and related phenomena were not studied. Here, the $A^2$ terms should be maintained in the system's Hamiltonian, and two-mode situation including different frequencies case is considered. It should be noted that the interaction Hamiltonian $\widehat{H}_I$ in Eq. (2) can be degenerated into single mode case when the indices $i = j$, that is the two photons have the same frequency.

The free Hamiltonian of electron and electromagnetic field are $-i\hbar v_e \nabla$ and $\sum_i \hbar \omega_i \hat{a}_i^\dagger \hat{a}_i$, respectively. Having the light field quantization be operated in macroscopic quantum electrodynamics (MQED) [35-36], the vector potential $\widehat{\boldsymbol{A}}$ for photonic mode $i$ is written as

$$\widehat{\boldsymbol{A}}_i(\boldsymbol{r}) = \sqrt{\frac{\hbar}{2\varepsilon_0 \omega_i}} \left[ \boldsymbol{F}_i(\boldsymbol{r}) \hat{a}_i + \boldsymbol{F}_i^*(\boldsymbol{r}) \hat{a}_i^\dagger \right] \tag{3}$$

where $\hat{a}_i$ and $\hat{a}_i^\dagger$ are the photon annihilation operator and creation operator of eigenmode with eigenfrequency $\omega_i$, respectively, satisfying commuter relationship $[\hat{a}_i, \hat{a}_i^\dagger] = 1$, and $\boldsymbol{F}_i(\boldsymbol{r})$ is the normalized mode profile of photonic mode with frequency $\omega_i$. Substituting Eq. (3) into Eq. (2), we can get the interaction Hamiltonian in interaction picture $\widehat{V}_I(t) = \widehat{V}_1(t) + \widehat{V}_2(t)$, where

$$\widehat{V}_1(t) = \sum_{i=1,2} \left[ W_i(z + v_e t) \hat{a}_i e^{-i\omega_i t} + W_i^*(z + v_e t) \hat{a}_i^\dagger e^{i\omega_i t} \right],$$

$$\widehat{V}_2(t) = \sum_{i,j=1,2} \frac{e^2}{2m_e \gamma} \frac{\hbar}{2\varepsilon_0 \sqrt{\omega_i \omega_j}} \Big[ \boldsymbol{F}'_i(z + v_e t) \boldsymbol{F}'_j(z + v_e t) \hat{a}_i \hat{a}_j e^{-i(\omega_i + \omega_j)t} +$$

$$\boldsymbol{F}'^*_i(z + v_e t) \boldsymbol{F}'^*_j(z + v_e t) \hat{a}_i^\dagger \hat{a}_j^\dagger e^{i(\omega_i + \omega_j)t} + \boldsymbol{F}'_i(z + v_e t) \boldsymbol{F}'^*_j(z + v_e t) \hat{a}_i \hat{a}_j^\dagger e^{-i(\omega_i - \omega_j)t}$$

$$+ \boldsymbol{F}'^*_i(z + v_e t) \boldsymbol{F}'_j(z + v_e t) \hat{a}_i^\dagger \hat{a}_j e^{i(\omega_i - \omega_j)t} \Big]. \tag{4}$$

Here $\boldsymbol{F}'_i(\boldsymbol{r}) = F_{ix}(\boldsymbol{r})\boldsymbol{e}_x + F_{iy}(\boldsymbol{r})\boldsymbol{e}_y + F_{iz}(\boldsymbol{r})\boldsymbol{e}_z/\gamma$. The more detailed algebraic derivation can be found in Supplementary Material section S1. In the interaction picture, the scattering operator can be written as [37-38]

$$\hat{S}(-\infty, \infty) = T\exp\left[-\frac{i}{\hbar}\int_{-\infty}^{\infty} dt \hat{V}_I(t)\right] \quad (5)$$

where $T$ denotes the time ordering. According to the scattering matrix description [4, 8], the relation between the initial state $|\psi_i\rangle$ and the final state $|\psi_f\rangle$ of joint quantum state of the system is connected by the scattering operator $\hat{S}$ as $|\psi_f\rangle = \hat{S}|\psi_i\rangle$.

The Eq. (5) can be expressed in terms of the Magnus expansion as $\hat{S}(-\infty, \infty) = \exp(\sum_{k=0}^{\infty} \tilde{S}_k)$, whose first term can be expressed as

$$\tilde{S}_1 = \sum_{i,j=1,2} \left[g_{qu,i}\hat{b}_i\hat{a}_i^\dagger - g_{qu,i}^*\hat{b}_i^\dagger\hat{a}_i + g_{qu,ij}^{(2)}\hat{b}_i^\dagger\hat{b}_j^\dagger\hat{a}_i\hat{a}_j - g_{qu,ij}^{(2)*}\hat{b}_i\hat{b}_j\hat{a}_i^\dagger\hat{a}_j^\dagger\right.$$

$$\left.-g_{p,ij}\hat{b}_i^\dagger\hat{b}_j\hat{a}_i\hat{a}_j^\dagger + g_{p,ij}^*\hat{b}_i\hat{b}_j^\dagger\hat{a}_i^\dagger\hat{a}_j\right] \quad (6)$$

where $\hat{b}_i = \exp(-i\omega_i z/v_e)$ is the electron lowering operators with ladder spacing $\hbar\omega_i$ and $\hat{b}_i^\dagger$ is its conjugate, satisfying commuter relation $[\hat{b}_i, \hat{b}_i^\dagger] = 0$. In Eq. (6), the $g_{qu,i}$ is the same as that defined in PINEM [8, 21]:

$$g_{qu,i} = \frac{e}{\hbar\omega_i}\int_{-\infty}^{\infty} dz' E_{i,z}(z')e^{-i\frac{\omega_i}{v_e}z'} \quad (7)$$

which is referred as the first-order quantum coupling constant in this paper. Similarly, here we define the second-order quantum coupling constants $g_{qu,ij}^{(2)}$ and $g_{p,ij}$ to represent the interaction strength between free electron and two-photons, which are given as

$$g_{qu,ij}^{(2)} = \frac{ie^2}{\hbar\omega_i\omega_j \cdot 2m_e\gamma}\frac{1}{v_e}\int_{-\infty}^{\infty} dz' \mathbf{E}_i'(z')\mathbf{E}_j'(z')\,e^{-i\frac{\omega_i+\omega_j}{v_e}z'}, \quad (8)$$

$$g_{p,ij} = \frac{ie^2}{\hbar\omega_i\omega_j \cdot 2m_e\gamma}\frac{1}{v_e}\int_{-\infty}^{\infty} dz' \mathbf{E}_i'(z')\mathbf{E}_j'^*(z')e^{-i\frac{\omega_i-\omega_j}{v_e}z'}, \quad (9)$$

where $\mathbf{E}_i'(\mathbf{r}) = E_{ix}(\mathbf{r})\mathbf{e}_x + E_{iy}(\mathbf{r})\mathbf{e}_y + E_{iz}(\mathbf{r})\mathbf{e}_z/\gamma$. The $g_{qu,ij}^{(2)}$ is related to emitting or absorbing two-photons, while $g_{p,ij}$ is related to absorbing one photon and emitting one photon simultaneously. It can be seen that $g_{qu,ij}^{(2)}$ and $g_{p,ij}$ are decided by the electric field components along different directions and the velocity of free electron.

To obtain the complete scattering operator, the higher order terms of Magnus expansion should also be calculated. The interaction Hamiltonian $\hat{V}_I(t)$ in interaction picture satisfies following commutation relation: $[\hat{V}_I(t), \hat{V}_I(t')] = [\hat{V}_1(t), \hat{V}_1(t')] +$ $[\hat{V}_1(t), \hat{V}_2(t')] + [\hat{V}_2(t), \hat{V}_1(t')] + [\hat{V}_2(t), \hat{V}_2(t')]$. Without loss of generality, for strong electron-light interaction and simplicity, we only consider the case satisfying phase matching condition, that is the phase velocity of light field $v_p$ equals to $v_e$. Therefore, it can be found $[\hat{V}_I(t), \hat{V}_I(t')] = constant$, $[\hat{V}_I(t''), [\hat{V}_I(t), \hat{V}_I(t')]] = 0$, and the scattering operator can ultimately be obtained as:

$$\hat{S} = e^{i\chi} e^{\sum_{i,j=1,2} \left[ g_{qu,i} \hat{b}_i \hat{a}_i^\dagger - g_{qu,i}^* \hat{b}_i^\dagger \hat{a}_i + g_{qu,ij}^{(2)} \hat{b}_i^\dagger \hat{b}_j^\dagger \hat{a}_i \hat{a}_j - g_{qu,ij}^{(2)*} \hat{b}_i \hat{b}_j \hat{a}_i^\dagger \hat{a}_j^\dagger - g_{p,ij} \hat{b}_i^\dagger \hat{b}_j \hat{a}_i \hat{a}_j^\dagger + g_{p,ij}^* \hat{b}_i \hat{b}_j^\dagger \hat{a}_i^\dagger \hat{a}_j \right]}$$

(10)

The scattering operator $\hat{S}$ in Eq. (10) is general to free electron-light interaction when considering two-photon process. As for single mode case with frequency $\omega_0$, the scattering operator $\hat{S}$ can be further reduced to

$$\hat{S} = e^{i\chi} e^{g_{qu} \hat{b}_0 \hat{a}^\dagger - g_{qu}^* \hat{b}_0^\dagger \hat{a} + g_{qu}^{(2)} \hat{b}_0^{\dagger 2} \hat{a}^2 - g_{qu}^{(2)*} \hat{b}_0^2 \hat{a}^{\dagger 2} - g_p (\hat{a} \hat{a}^\dagger + \hat{a}^\dagger \hat{a})}$$

(11)

where $\chi$ is the global phase as discussed in Ref. [39], while $g_{qu} = g_{qu,0}$, $g_{qu}^{(2)} = g_{qu,00}^{(2)}$, $g_p = g_{p,00}$ and $g_p = i|g_{qu}^{(2)}|$ under phase matching condition. More derivation details can be found in Supplementary Material section S1. The $\hat{S}$ in Eq. (11) can be further split into a product of four exponentials:

$$\hat{S} = e^{i\chi} e^{g_{qu} \hat{b}_0 \hat{a}^\dagger - g_{qu}^* \hat{b}_0^\dagger \hat{a}} e^{g_{qu}^{(2)'} \hat{b}_0^{\dagger 2} \hat{a}^2 - g_{qu}^{(2)'*} \hat{b}_0^2 \hat{a}^{\dagger 2}} e^{-g_p' (\hat{a} \hat{a}^\dagger + \hat{a}^\dagger \hat{a})}$$

(12)

The scattering operator $\hat{S}$ in Eq. (10) ~ Eq. (12) not only are closely related to previous reported results but also include more phenomena of free electron-light interaction.

(1) It is found that the scattering operator in Eq. (12) has similar form with that in Ref. [34], proving our universal results can degenerate to existing theoretical results. From Eq. (12), it is also noticed that the term $e^{g_{qu} \hat{b}_0 \hat{a}^\dagger - g_{qu}^* \hat{b}_0^\dagger \hat{a}}$ in $\hat{S}$ is the same as the scattering operator in Ref. [6-7], which represents single-photon process in electron-light interaction. The result reveals that the first-order and second-order quantum coupling constants $g_{qu}$ and $g_{qu}^{(2)}$ are independent to each other, showing the same behavior as the nonlinear susceptibility in nonlinear optics.

(2) The $g_{qu}^{(2)}$ and $g_p$ reveal the strength of two-photon process and, for photonic modes in previous reports [8, 23-24, 40-44], their value are several orders smaller than $g_{qu}$ according to the calculation in the Section III-part 1. Thus Eq. (11) and Eq. (12) degenerate to the form just considering single-photon process. Moreover, how to enhance and manipulate $g_{qu}^{(2)}$ will be given in Section III-part 1.

(3) The term $\exp(g_{qu}^{(2)'}\hat{b}_0^{\dagger 2}\hat{a}^2 - g_{qu}^{(2)'*}\hat{b}_0^2\hat{a}^{\dagger 2})$ in $\hat{S}$ represents two-photon emission and/or absorption process. The second-order quantum coupling constant $g_{qu}^{(2)}$ is modified as $g_{qu}^{(2)'}$ in this term. The relationship between $g_{qu}^{(2)'}$ and $g_{qu}^{(2)}$ given in Supplementary Material section S1 shows that the deviation of $\left|g_{qu}^{(2)'}\right|$ from $\left|g_{qu}^{(2)}\right|$ can be ignored when $\left|g_{qu}^{(2)}\right|$ is small (e.g., $\left|g_{qu}^{(2)}\right|<0.2$).

(4) The $\hat{S}$ in Eq. (10) will degenerate into Eq. (12) for single mode situation. The term $\exp[-g_p'(\hat{a}\hat{a}^\dagger + \hat{a}^\dagger\hat{a})]$ in Eq. (12) represents the process of absorbing one photon and emitting one photon simultaneously, and the concrete expression of $g_p'$ is also given in Supplementary Material section S1. Although the total photon number and the electron energy are not changed, a phase is brought into the joint state and the electron as well as photonic pattern might be modified. The KD effect can be well described by the term $\exp[-g_p'(\hat{a}\hat{a}^\dagger + \hat{a}^\dagger\hat{a})]$, and the corresponding probability of electron's transverse momentum is deduced and discussed in details in Section III-part 4.

(5) Besides the elastic scattering of free electrons by standing wave in KD effect, the $[-g_{p,ij}\hat{b}_i^\dagger \hat{b}_j \hat{a}_i \hat{a}_j^\dagger + g_{p,ij}^*\hat{b}_i \hat{b}_j^\dagger \hat{a}_i^\dagger \hat{a}_j]$ term in Eq. (10) can be used to explain the nonlinear Compton scattering, which is the inelastic scattering of free electrons by traveling wave field. In Section III-part 5, the experiment results of nonlinear Compton scattering are well fitted based on our derived $\hat{S}$, and the phenomenon of free electrons induced entanglement between two photonic states is predicted.

**The joint density matrix of free electron and light**

In this part, the joint density matrix of free electron and photons, namely $\rho_{tot}^f = \hat{S}\rho_{tot}^i\hat{S}^\dagger$, is deduced after obtaining the $\hat{S}$ for specific situations.

(1) First, the single mode case is considered by assuming the incident electron beam energy as $E_0$ (initial electron state $|E_0\rangle$) and no photons in the system (vacuum state $|0\rangle$), so that the initial joint density matrix is $\rho_{tot}^i = |E_0\rangle\langle E_0|\otimes|0\rangle\langle 0|$. With the scattering operator given in Eq. (12), the full density matrix of final state can be written as $\rho_{tot}^f = |\Psi_{e-p}\rangle\langle\Psi_{e-p}|$ with the electron-photon state expressed as $|\Psi_{e-p}\rangle = \sum_{p=0}^{\infty} C_p^0|E_0 - p\hbar\omega_0\rangle|p\rangle$. Without considering the phase terms of $\hat{S}$, the coefficient $C_p^0$ is

$$C_p^0 = \frac{1}{\sqrt{\cosh 2\left|g_{qu}^{(2)'}\right|}} e^{-\frac{|g_{qu}|^2}{2}} e^{ip\varphi_{g1}} \sum_{l=\max(0,p-2m)}^{\infty} \sum_{m=0}^{\infty} \frac{(-1)^{l+3m-p}|g_{qu}|^{2l+2m-p}}{(l+2m-p)!\,l!\,m!}$$

$$\times \frac{e^{-im\Delta\varphi_g}\left(\frac{1}{2}\tanh 2\left|g_{qu}^{(2)'}\right|\right)^m (2m+l)!}{\sqrt{p!}} \quad (13)$$

where $\Delta\varphi_g = 2\varphi_{g1} + \varphi_{g2}$ is the phase between single-photon process and two-photon process, with $\varphi_{g2} = \arg(g_{qu}^{(2)'})$ and $\varphi_{g1} = \arg(g_{qu})$ (more derivation details can be found in Supplementary Material section S2). In Eq. (13), an abnormal phase term $e^{-im\Delta\varphi_g}$ appears and could affect the final state obviously. The relate phenomenon will be discussed in Section III-part 3.

According to Ref. [8], the diagonal elements of $\rho_{tot}^f$ is $P_{nk} = (\langle n|\otimes\langle E_0 + k\hbar\omega_0|)\rho_{tot}^f(|E_0 + k\hbar\omega_0\rangle\otimes|n\rangle)$ after electron interacting with photons. Actually, $P_{nk}$ represents the coincident probability for the photon number state $|n\rangle$ with the electron energy gain $k\hbar\omega_0$ [6], and the energy spectrum of free electrons after interaction is $P_k = \sum_n P_{nk}$. The calculation results of $P_{nk}$ and $P_k$ for free electron emitting photons without ignoring two-photon process are shown in Section III -part 2. On the other hand, the electron density matrix can also be obtained from full density matrix as

$$\rho_{el}^f = \text{Tr}_{ph}(\rho_{tot}^f) = \sum_{k=-\infty}^{\infty}\sum_{k'=-\infty}^{\infty} |C_p^0|^2 |E_0 + k\hbar\omega_0\rangle\langle E_0 + k'\hbar\omega_0|\delta_{kk'}\delta_{n,-k} \quad (14)$$

Accordingly, the entanglement between free electron and its emitted photons can be deduced from the electron density matrix, which will also be discussed carefully in Section III-part 2.

(2) As for the situation that incident photonic state is a coherent state $|\alpha\rangle$ and initial electron state is $|E_0\rangle$, the full density matrix of final electron-photon state can be written as $\rho_{tot}^f = |\Psi_{e-p}\rangle\langle\Psi_{e-p}|$ with $|\Psi_{e-p}\rangle = \sum_{p=-\infty}^{\infty} C_p^n |E_0 - p\hbar\omega_0\rangle|n+p\rangle$, where the concrete expression of the coefficients are

$$C_p^n = e^{\frac{|g_{qu}|^2}{2}} e^{-g_p'(2n+1)} e^{ip\varphi_{g1}} \sum_{l=\max(0,p-2m)}^{\infty}\sum_{m=-q}^{\infty}\sum_{q=0}^{n/2} \frac{(-1)^{l-p+q+3m}|g_{qu}|^{2l-p+2m}}{(l-p+2m)!\,l!}$$

$$\times \frac{\left(\frac{1}{2}\tanh 2\left|g_{qu}^{(2)'}\right|\right)^{2q+m} e^{-im\Delta\varphi_g}}{(q+m)!\,q!} \left(\cosh 2\left|g_{qu}^{(2)'}\right|\right)^{-\frac{1}{2}[2(n-2q)+1]} \sqrt{\frac{n!}{(n+p)!}\frac{(n+2m+l)!}{(n-2q)!}}$$

(15)

The detailed derivation of Eq. (15) is given in Supplementary Material Section S2.4. It is worth noting that, for the strong coherent state ($\alpha \gg 1$) and weak interaction strength ($|g_{qu}| \ll 1$, $\left|g_{qu}^{(2)'}\right| \ll 1$), the coefficients in Eq. (15) can be further simplified as

$$C_p^n \approx e^{-g_p'(2n+1)}e^{ip\varphi_{g1}} \sum_{m=-\infty}^{\infty} J_{p-2m}(2|g_1|)J_{-m}(2|g_2|)e^{-im\Delta\varphi_g} \qquad (16)$$

where the coupling constants are $g_1 = g_{qu}\sqrt{\bar{n}}$, $g_2 = g_{qu}^{(2)}\bar{n}$, $\bar{n}$ is the average number of photons. When $g_{qu}^{(2)} = 0$ or $g_{qu} = 0$, the coefficient given in Eq. (16) degenerates into $C_p^n = e^{ip\varphi_{g1}}J_p(2|g_1|)$ and $C_p^n = e^{-g_p'(2n+1)}e^{-ip\varphi_{g2}/2}J_{-p/2}(2|g_2|)$, respectively. When there is external incident light, the related phenomena and further discussion of the final electron-photon states are given in Section III-part 3 and Supplementary Material Section S2.5. With Eq. (16), it is found that the free electron energy spectrum after interaction changes with the $\Delta\varphi_g$ and becomes asymmetric.

## III. Calculation results and discussion

Based on the theory in Section II, the novel and existing phenomena caused by the electron-two-photon interaction are discussed in the five parts of this Section. Part 1 discusses how to enhance two-photon process through metasurface; Part 2 gives the calculated final electron-photon joint state and related entanglement properties of electron energy state and photon number state; Part 3 reveals the quantum interference between single-photon process and two-photon process; In part 4 and part 5, the Kapitza-Dirac effect and nonlinear Compton scattering are explained by the analytical full quantum theory based on $\hat{S}$, respectively.

### 1. Manipulate the first-order & second-order quantum coupling constants

In conventional PINEM, microsphere and ring cavity [24, 40-41], photonic crystal [42-44], dielectric laser accelerator [8, 45] and metal-dielectric multilayer structure [23] were used to increase quantum coupling constant $g_{qu}$, achieving enhancement of electron-light interaction (e.g., $g_{qu}$ had reached $\sim 1$ in experiments [23]). According to Eq. (7) ~ Eq. (9) and the near field components in the previous reports, the second-order quantum coupling constant $g_{qu}^{(2)}$ and $g_p$ defined in this papers are calculated and given in Table I, which are all less than $10^{-8}$ and nine~ten orders of magnitude smaller than $g_{qu}$. Thus, the scattering operator shown in Eq. (12) degrades into the classic form $\hat{S} = e^{i\chi}e^{g_{qu}\hat{b}_0\hat{a}^\dagger - g_{qu}^*\hat{b}_0^\dagger\hat{a}}$ in PINEM only considering single-photon process [6-8], and the two-photon process can be ignored in previous experiments. Nevertheless, to make the two-photon process obvious, the Eq. (8) and Eq. (9) in Section II indicate that strong $g_{qu}^{(2)}$ and $g_p$ could be realized by enhancing modulus of the electric field $|E|$ and meanwhile relatively reducing $z$-component of the electric field $E_z$ for suppressing single-photon process (as $g_{qu}$ is decided by $E_z$).

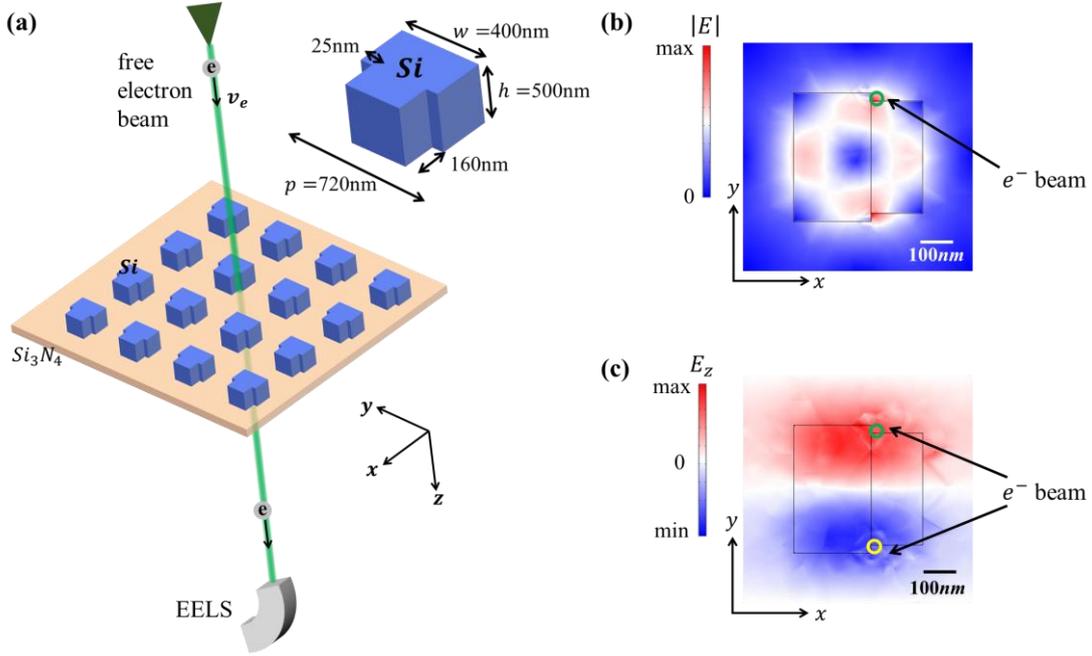

**Fig. 2. The scheme of enhancing $\left|g_{qu}^{(2)}\right|$ and changing $\Delta\varphi_g$ based on silicon quasi bound state in the continuum (BIC) metasurface. (a)** The free electron beam interacts with the quasi-BIC metasurface enhancing the two-photon process and is detected by electron energy loss spectroscopy (EELS). The parameters of the metasurface are given in the inset. **(b)** The simulated electric field distribution $|E|$ of the quasi-BIC mode, and the green circle denotes the position where electron beam incidents in. **(c)** The simulated $E_z$ of the quasi-BIC mode, and the green and yellow circle mark the different electron incident positions, which have difference $\Delta\varphi_g$ between $g_{qu}$ and $g_{qu}^{(2)'}$.

As shown in Fig. 2(a), we propose to have free electron beam pass through the metasurface with photonic bound state in the continuum (BIC) [46-47], which has strong transverse component of electric field to relatively enhance $g_{qu}^{(2)}$ and $g_p$. The silicon metasurface structure with a thin layer of silicon nitride substrate is similar to that in Ref. [48] and the parameters showing in the inset of Fig. 2(a) are designed to have relative strong transverse electric field. The quasi-BIC mode is simulated using finite element method and the electric field distribution $|E|$ is shown in Fig. 2(b), where the electric field is greatly enhanced at the corners. If the 200keV electron beam is incident along $z$-axis at the green circle, the $\left|g_{qu}^{(2)}\right|$ can reach up to ~0.099 and the $|g_{qu}|$ is only 0.041. The comparison of $|g_{qu}|$ with $\left|g_{qu}^{(2)}\right|$ and $g_p$ for different photonic structures is given in Table I, which indicates that the $\left|g_{qu}^{(2)}\right|$ and $g_p$ have been enhanced by more than eight orders of magnitude by applying the quasi-BIC mode and even larger than $g_{qu}$.

**TABLE I.** Comparison of quantum coupling constant $g_{qu}$, $g_{qu}^{(2)}$ and $g_p$ for different photonic modes

| Photonic Structure | | $|g_{qu}|$ | $|g_{qu}^{(2)}|$ | $|g_p|$ |
|---|---|---|---|---|
| Prism [5] | 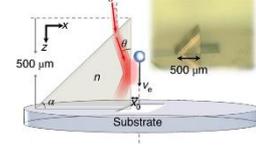 | 0.0028 | 7.2×10⁻¹⁵ | 7.2×10⁻¹⁵ |
| Microsphere cavity [40] | 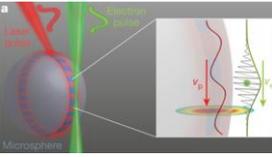 | 0.0008 | 3.2×10⁻¹⁴ | 3.2×10⁻¹⁴ |
| Photonic crystal [43] | 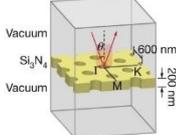 | 0.0004 | 8.0×10⁻¹⁴ | 8.0×10⁻¹⁴ |
| Dielectric laser accelerator [8] | 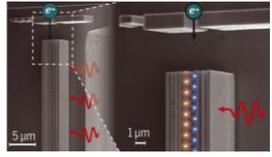 | 0.016 | 5.0×10⁻¹³ | 5.0×10⁻¹³ |
| Ring cavity [41] | 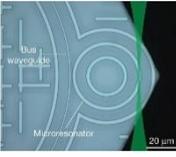 | 0.03 | 3.9×10⁻¹² | 3.9×10⁻¹² |
| Multilayer structure [23] | 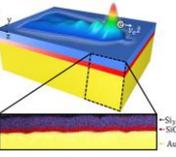 | 0.99 | 1.7×10⁻⁹ | 1.7×10⁻⁹ |
| **Quasi-BIC in this work** | 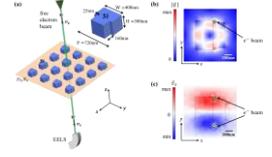 | 0.041 | 0.099 | 0.099 |

Note: The second order quantum coupling constant $g_{qu}^{(2)}$ and $g_p$ are not given in previous work but calculated according to Eq. (8) and Eq. (9).

With $|g_{qu}^{(2)}| \approx 0.099$, $|g_{qu}| \approx 0$ and $\Delta\varphi_g = 0$, the probability of emitting a pair of photons simultaneously is $|C_p^0|^2 \approx 2\%$ according to Eq. (13). Considering the single-photon emission with an intrinsic generation probability of ~2.5% had been observed [24], the scheme proposed here should be feasible for observing two-photon emission by detecting the electron energy loss spectrum in the experiment. And more calculation results and analysis of photon emission by free electron under different quantum coupling constants are given in part 2.

In addition, the phase $\Delta\varphi_g = 2\varphi_{g1} + \varphi_{g2}$ related to $g_{qu}$ and $g_{qu}^{(2)'}$ plays an important role in emitting photons by free electron, especially the quantum interference between single-photon and two-photon processes as discussed in part 3. The $\Delta\varphi_g$ can be varied by changing the incident position of free electron beam. As the electron beam pointing at the symmetric positions in Fig. 2(c), we have different $\varphi_{g1}$ but the same $\varphi_{g2}$ for green circle and yellow circle. Therefore, $\Delta\varphi_g$ can be regulating by changing the incident position of the electron beam in this proposed scheme, which is feasible for ultrafast transmission electron microscopes (UTEM).

## 2. The final states of joint quantum state and the entanglement between free electron and photons

In this part, when considering the initial photonic state is a vacuum state, the final states of free electron and emitted photons are studied and analyzed to further show the two-photon process, and meanwhile the free electron-photons entanglements are given quantitatively. Firstly, we calculate the diagonal elements of full density matrix of the joint quantum state according to Eq. (13). It not only provides the experimentally measurable quantities such as energy spectrum of free electrons and the photon number states, but also indicates the corresponding electron-photon entanglement patterns. Since the $g_{qu}$ and $g_{qu}^{(2)}$ could be manipulated by designing the photonic structures and their field components, as revealed by the results and analysis about Fig. 2, the modulus and argument of $g_{qu}$ and $g_{qu}^{(2)}$ are set reasonably in the following calculations to find out the different states of free electron and emitted photons.

Considering a special case of $\Delta\varphi_g = 0$, e.g., $\varphi_{g1} = \varphi_{g2} = 0$, the color map of the coincident probabilities $P_{nk}$ for the photon number state $|n\rangle$ with free electron gaining energy $k\hbar\omega_0$ can be calculated according to Eq. (13) in Section II. Further, the probability $P_k$ of free electron energy gain $k\hbar\omega_0$, namely the electron energy spectrums, can be obtained from $P_{nk}$. The following are three cases with different $g_{qu}$ and $g_{qu}^{(2)}$:

(1) Assuming $|g_{qu}| = 0.04$ and $|g_{qu}^{(2)}| = 0.1$, the calculated $P_{nk}$ and $P_k$ are shown in Fig. 3(a) and Fig. 3(d), respectively. In this case, the single-photon emission is weak and the two-photon emission is allowed. Free electrons only emit even number of photons, e.g., two photons. The corresponding probability of emitting two photons by an electron is ~2% in this case, which can be detected in the experiment using the UTEM for single-photon emission with probability of ~2.5% in Ref. [24]. In the proposed scheme using silicon quasi-BIC metasurface in Fig. 2, the $|g_{qu}^{(2)}|$ can reach

~0.1 and $|g_{qu}|$ is only ~0.04, which meets the assumption of much larger $|g_{qu}^{(2)}|$ than $|g_{qu}|$. To achieve more obvious two-photon emission, $|g_{qu}^{(2)}|$ should be further increased, which indicates stronger free electron-two-photon interaction.

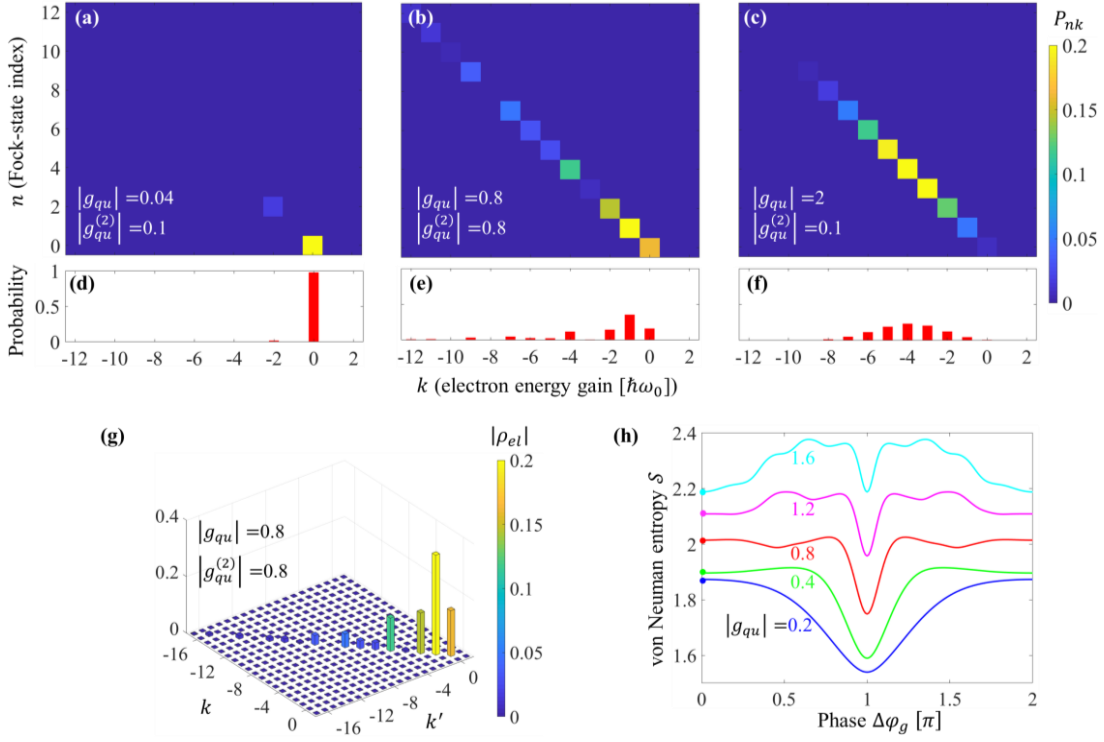

**Fig. 3. The diagonal elements of total density matrix $P_{nk}$ after interaction and the entanglement between electron and photons. (a)-(f)** The calculated diagonal elements of total density matrix $P_{nk}$ (shown in (a), (b) and (c)) and the electron energy probability $P_k$ ((d), (e) and (f), red bars) after electron emitting photons. The corresponding parameters are assumed $g_{qu} =0$, $g_{qu}^{(2)} =0.1$ in (a) and (d), $g_{qu} =0.8$, $g_{qu}^{(2)} =0.8$ in (b) and (e), $g_{qu} =2$, $g_{qu}^{(2)} =0.1$ in (c) and (f). **(g)** The calculated electron density matrix after emission. There exist only diagonal elements implying that electron and photons are entangled when assuming $|g_{qu}| = |g_{qu}^{(2)}| =0.8$. The phase $\Delta\varphi_g$ in (a)-(g) are all set as 0. **(h)** The von Neuman entanglement entropy $S$ of electron-photons as a function of $\Delta\varphi_g$. $|g_{qu}^{(2)}|$ of all lines are set as 0.8, while $|g_{qu}|$ are varied as 0.2, 0.4, 0.8, 1.2 and 1.6.

(2) Assuming $|g_{qu}| =0.8$ and $|g_{qu}^{(2)}| =0.8$, the $P_{nk}$ and $P_k$ are given in Fig. 3(b) and Fig. 3(e), respectively. It can be found that the probabilities of emitting two photons and four photons are ~14.5% and ~11.9%, respectively, revealing the significant interaction strength between free electron and two-photons. The electron energy spectrum $P_k$ neither is a Poisson distribution corresponding to single-photon process, nor only emit even number of photons. Meanwhile, we notice that the probability of losing $3\hbar\omega_0$ for an electron is only ~0.76% when $|g_{qu}|$ (single-photon emission) and $|g_{qu}^{(2)}|$ (two-photon emission) are comparable, which is beyond common knowledge. Actually, the abnormal joint state is attributed to the quantum interference between single-photon process and two-photon processes, which will be further discussed in part 3.

(3) Assuming $|g_{qu}|$ =2 and $|g_{qu}^{(2)}|$ =0.1, the $P_{nk}$ and $P_k$ are shown in Fig. 3(c) and Fig. 3(f), respectively. In this case, the single-photon emission is dominant and the strength of two-photon emission is weaker due to $|g_{qu}| \gg |g_{qu}^{(2)}|$. The electron energy spectrum $P_k$ has a nearly Poisson distribution, which is the feature of single-photon PINEM. Actually, $|g_{qu}^{(2)}|=0.1$ means there still exist a considerable two-photon process, which causes the $P_k$ in Fig. 3(f) slightly deviating from the Poisson distribution. And the $P_k$ will gradually degenerate into the single-photon PINEM when $|g_{qu}^{(2)}|$ keeps decreasing to 0.

In the above situations, the free electron and the emitted photons are correlated, which is indicated by the electron density matrix or the photon density matrix. According to Eq. (14), the electron density matrix is calculated and illustrated in Fig. 3(g), which has only diagonal elements by choosing quantum coupling constants as $|g_{qu}| = |g_{qu}^{(2)}|$ =0.8 for more obvious correlation features. The purity of the electron density matrix (Purity = $\text{Tr}(\rho_{el}^2)/(\text{Tr}\rho_{el})^2 = \sum_k P_k^2$ [8]) is also estimated according to Fig. 3(g). The Purity value of 0.193 is much smaller than 1 indicating that the final electron-photon quantum state is a mixed state, that is the electron state and photon state are not separable.

To quantify the degree of entanglement quantitatively, the von Neuman entropy $S$ is used to reveal the entanglement between free electron and emitted photons. The $S$ is defined as $S = -\text{Tr}[\rho_{ph}\ln(\rho_{ph})] = -\text{Tr}[\rho_{el}\ln(\rho_{el})]$ [49], where $\rho_{ph}$ and $\rho_{el}$ are the photon density matrix and electron density matrix, respectively. From the $\rho_{el}$ in Fig. 3(g), we can obtain the corresponding $S$ =2.01. It is larger than the entropy of Bell states $S = \ln(2)$, the maximum entangled states for a two-qubit system. The large $S$ indicates the strong entanglement between free electron energy state and photon number state. When fixing $|g_{qu}^{(2)}|$ =0.8 and $\Delta\varphi_g = 0$, the $S$ increases with $g_{qu}$, as illustrated by the dots of different colors in Fig. 3(h). Among these dots, even the smallest value $S = 1.87$ (when $|g_{qu}| = 0.2$ and $|g_{qu}^{(2)}| = 0.8$) is larger than the maximal $S$ =1.30 with $|g_{qu}| = 1$ in single-photon PINEM [23], which reveals the increased electron-photon entanglement due to two-photon process.

Besides, we also find that the $S$ changes drastically when regulating $\Delta\varphi_g$, which is illustrated by the lines in Fig. 3(h) with fixed $|g_{qu}^{(2)}| = 0.8$ and different $|g_{qu}|$. It can be found $S$ is symmetric with respect to $\Delta\varphi_g = \pi$. And the minimum value of $S$ at

$\Delta\varphi_g = \pi$, namely the weakest electron-photons correlation, may origin from the interference between single-photon process and two-photon process, as discussed in part 3. Comparing different lines, it is clearly $\mathcal{S}$ enhances with increased $|g_{qu}|$ for any $\Delta\varphi_g$. However, the maximum $\mathcal{S}_{max}$ are acquired at different specific $\Delta\varphi_g$ for different lines, e.g., $\mathcal{S}_{max} = 1.8731$ at $\Delta\varphi_g = 0$ when $|g_{qu}| = 0.2$, while $\mathcal{S}_{max} = 2.3768$ at $\Delta\varphi_g = 0.65\pi$ when $|g_{qu}| = 1.6$. The variation characteristics of $\mathcal{S}$ with $\Delta\varphi_g$ become more complex as $|g_{qu}|$ increases.

## 3. The quantum interference between single-photon process and two-photon process

In this part, we focus on how $\Delta\varphi_g$ influences on the final electron-photon joint state. Firstly, considering the initial optical vacuum state of the system and according to the phase term $e^{-im\Delta\varphi_g}$ in Eq. (13), the electron energy spectrum varying with $\Delta\varphi_g$ is illustrated in Fig. 4 in form of color map. Under the following three conditions, the results reveal the features of interference fringes, which unveils the quantum interference between single-photon process and two-photon process in electron-light interaction.

(1) The $|g_{qu}|$ of single-photon process is larger than $|g_{qu}^{(2)}|$ of two-photon process, but there still exists a mild-strong two-photon process (setting $|g_{qu}| = 2$ and $|g_{qu}^{(2)}| = 0.2$). The top figure of Fig. 4(a) illustrates the electron energy spectrum as a function of $\Delta\varphi_g$, which is resembling interference fringes. The bottom figure reveals obviously different electron energy spectrums with $\Delta\varphi_g = \pi$ and $\Delta\varphi_g = 0$, and illustrates that the $\Delta\varphi_g = \pi$ corresponds to a broadened spectrum. This result reveals obvious quantum interference phenomenon between single-photon process and two-photon process.

As a comparison, when $|g_{qu}^{(2)}|$ is set as 0 (namely the two-photon process is ignorable), the energy spectrum of electron no longer changes with $\Delta\varphi_g$ as shown in Fig. 4(b). That is why the experimental result of single-photon PINEM [6] can not find this interference phenomenon. Besides, the calculated electron energy spectrum in Fig. 4(b) is exactly the same as single-photon PINEM theory with probability distribution of $P_k = e^{-|g_{qu}|^2}|g_{qu}|^{2k}/k!$ [6], which is natural because the analytical expression can also be obtained by simplifying Eq. (13) when $g_{qu}^{(2)} = 0$ (see more details in Supplementary Material section S2).

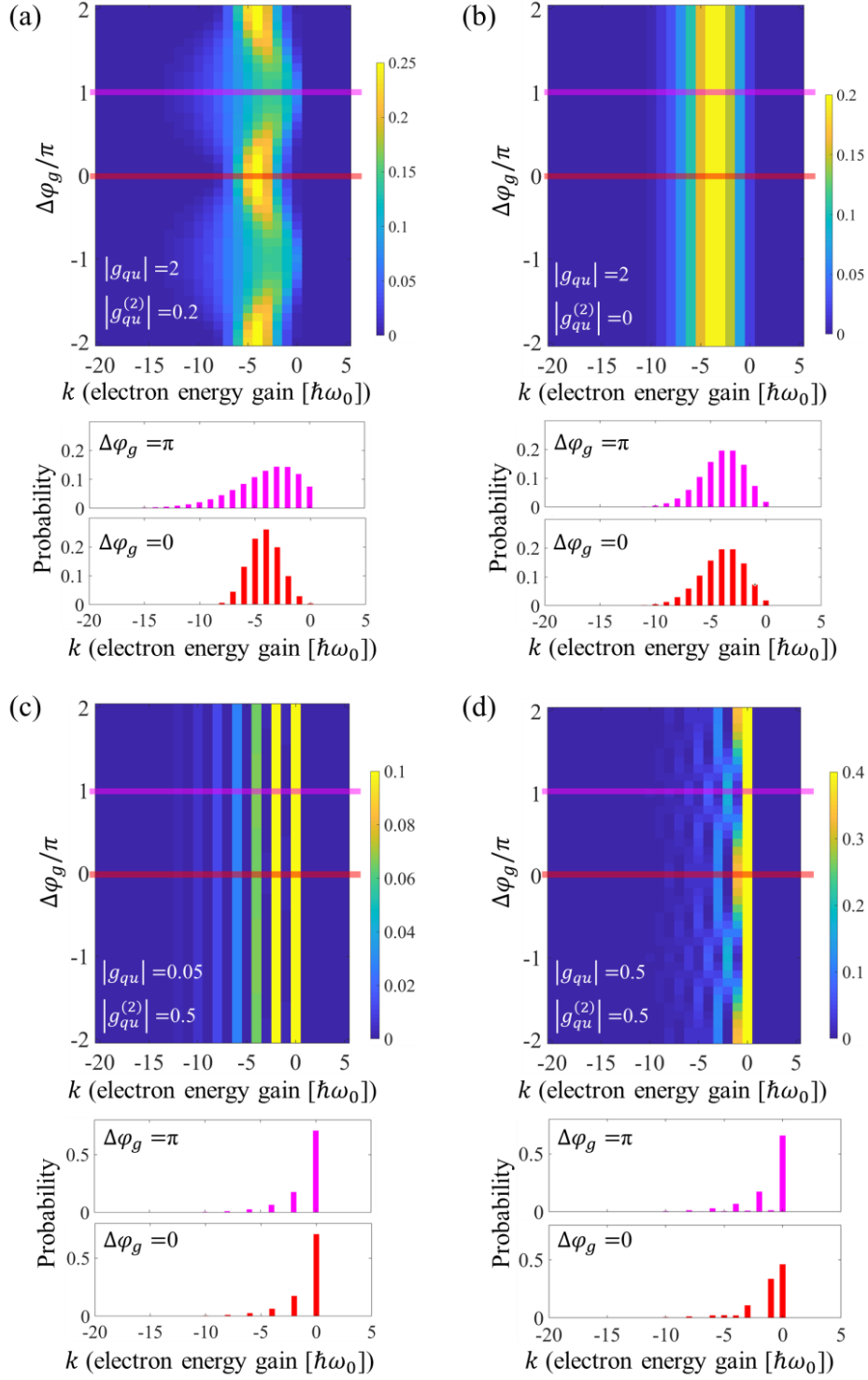

**Fig. 4. The electron energy spectrum varies with the phase $\Delta\varphi_g$ between first-order quantum coupling constant $g_{qu}$ and modified second-order quantum coupling constant $g_{qu}^{(2)'}$.** (a) The electron energy spectrum (EELS) after electron emitting photons varies with $\Delta\varphi_g$ is shown in form of color map in the top inset, the colors represent probability. The bar chart below is the corresponding EELS when $\Delta\varphi_g = \pi$ (denotes by magenta bars) and $\Delta\varphi_g = 0$ (denotes by red bars), respectively. The phase $\Delta\varphi_g$ varies from $-2\pi$ to $2\pi$, and the amplitude of quantum coupling constant are set as $|g_{qu}|=2$ and $|g_{qu}^{(2)}|=0.2$. (b)-(d) The same illustration as (a) but with (b) $|g_{qu}|=2$ and $|g_{qu}^{(2)}|=0$, (c) $|g_{qu}|=2$ and $|g_{qu}^{(2)}|=0$, (d) $|g_{qu}|=2$ and $|g_{qu}^{(2)}|=0$.

(2) The $|g_{qu}|$ of single-photon process is much smaller than $\left|g_{qu}^{(2)}\right|$ of two-photon process, but there still exists non-negligible $|g_{qu}|$ (setting $|g_{qu}| = 0.05$ and $\left|g_{qu}^{(2)}\right| = 0.5$). Fig. 4(c) reveals that the energy spectrum of free electron does not vary with $\Delta\varphi_g$, and free electron only loss energy of $2\hbar\omega_0$ and its integral multiples. Though the two-photon process is dominant in this case, there is no quantum interference like the first situation.

The reason of existing only stronger two-photon emission can be understood by taking a look at the interaction strength between free electron and light. Without considering the quantum interference effect, the interaction strength is $\left|g_{qu}^{(2)}\right|\bar{n}$ for two-photon emission but $|g_{qu}|\sqrt{\bar{n}}$ for single-photon emission, where $\bar{n}$ is the average number of photons (see more details in Supplementary Material section S2.4). Thus, the larger $\left|g_{qu}^{(2)}\right|$ and proportion to $\bar{n}$ result in the dominant two-photon process and vanish of interference phenomenon in this case.

(3) The value of $|g_{qu}|$ and $\left|g_{qu}^{(2)}\right|$ are comparable (setting $|g_{qu}| = 0.5$ and $\left|g_{qu}^{(2)}\right| = 0.5$). The calculated result in Fig. 4(d) (top figure) illustrates that there still exist fluctuating streaks in this case, revealing the variation of electron's energy spectrum with $\Delta\varphi_g$. The most unexpected discovery is that when $\varphi_g = \pi$ the EELS of free electron is resembled to that in Fig. 4(c), which means the single-photon process is suppressed by the quantum interference although the $|g_{qu}|$ is close to $\left|g_{qu}^{(2)}\right|$ (here $|g_{qu}|$ is picked as a considerable value). Further increasing $|g_{qu}|$ and $\left|g_{qu}^{(2)}\right|$ synchronously, it can be found that nearly only two-photon emission can be generated by free electrons even $|g_{qu}|$ is relatively large (see Supplementary Material section S2)!

As for the situation that photonic structure is pumped by an intense laser, comparing Eq. (16) and Eq. (13), it can be found the final electron-photon joint state also varies with the $\Delta\varphi_g$ only when single-photon or two-photon process is missing. Therefore, the quantum interference between two processes remains when there is external incident light. The electron energy spectrum after interaction is asymmetric due to the quantum interference phenomena, but becomes symmetric when $\Delta\varphi_g$ is an odd multiple of $\pi/2$. The asymmetric electron energy spectrums are very different from the traditional single-photon PINEM, and the related calculated results and further discussion are shown in Supplementary Material Section S2.5.

Similar quantum interference had also emerged between free electron-photon interaction (single-photon process with two light beams of different frequencies in traditional PINEM) [50-51] and free electron-atoms interaction [52]. However, what we discovery here is the quantum interference between first-order and second-order process of free electron-photon interaction. Due to the electron energy spectrum is directly related to the photon emission, this discovery suggests the photonic quantum state can be shaped by adjusting $\Delta\varphi_g$.

Moreover, the quantum interference also provides a theoretical support for directly detecting the full vector field of light, in particular the phase of field, using free electrons. Based on traditional PINEM research, only the longitudinal electric field component can be obtained directly. And the latest researches relied on reference of lights and algorithmic recovery to realize the full vector detection of light field based on a concept called 'Free-Electron Ramsey-Type Interferometry' [53-55]. The quantum interference discovered here theoretically provides a way of detecting each electric field component, as well as phase, of the light directly by free electrons.

4. **Two-photon process for the Kapitza-Dirac effect**

In Eq. (12), the term $e^{-g'_p(\hat{a}\hat{a}^\dagger+\hat{a}^\dagger\hat{a})}$ of the scattering operator $\hat{S}$ only introduces a global phase into the joint state and has no influence on the above calculation results about two-photon generation and quantum interference. However, this term is related to some effects revealing that free electron absorbs a photon and emits a photon simultaneously. This part indicates that the KD effect is related to $e^{-g'_p(\hat{a}\hat{a}^\dagger+\hat{a}^\dagger\hat{a})}$ and the electron diffraction pattern could be deduced by this term.

In the KD effect, free electrons pass through the standing light wave and are scattered by photons forming the interference pattern [28-30]. Accordingly, as depicted in Fig. 5(a), we consider that a plane free electron wave flies through a standing light wave formed by two counter incident light waves with the same frequency $\omega_0$ and wavevector $k_p \boldsymbol{e}_x$ and $-k_p \boldsymbol{e}_x$ ($k_p = \omega_0/c$). The electric field of the standing light wave can be written as $\boldsymbol{E}_s(x,t) = 2|E_0|\cos(k_p x)\cos(\omega_0 t)\boldsymbol{e}_y$. Since the higher-order terms of the Magnus expansion are generally ignored for the standing light wave changing with time [56], we only reserve the first term and obtain the Eq. (6), although the phase-matching condition is not met between electron and standing light wave. Thus, the scattering operator $\hat{S}$ is also written as Eq. (11).

It can be found that $g_{qu}$ and $g_{qu}^{(2)}$ are both negligible for this case, while only the $g_p$ is reserved, so the Eq. (11) is simplified as $\hat{S} = e^{-g_p(\hat{a}\hat{a}^\dagger+\hat{a}^\dagger\hat{a})}$. After careful mathematical derivation (see Supplementary Material section S3), we can get the wavefunction of free electron as

$$\psi_e(\mathbf{r},t) = \left[ e^{i\chi} e^{-i\frac{e^2|E_0|^2 L}{2\gamma m_e \hbar \omega_0^2 v_e}} \sum_{n=-\infty}^{\infty} (-i)^n J_n\left(\frac{e^2|E_0|^2 L}{2\gamma m_e \hbar \omega^2 v_e}\right) e^{in2k_p x} \right] \psi_0(\mathbf{r},t) \quad (17)$$

where $\psi_0(\mathbf{r},t)$ is the initial wavefunction of free electron, and $L$ is the interaction length between free electron and standing light wave.

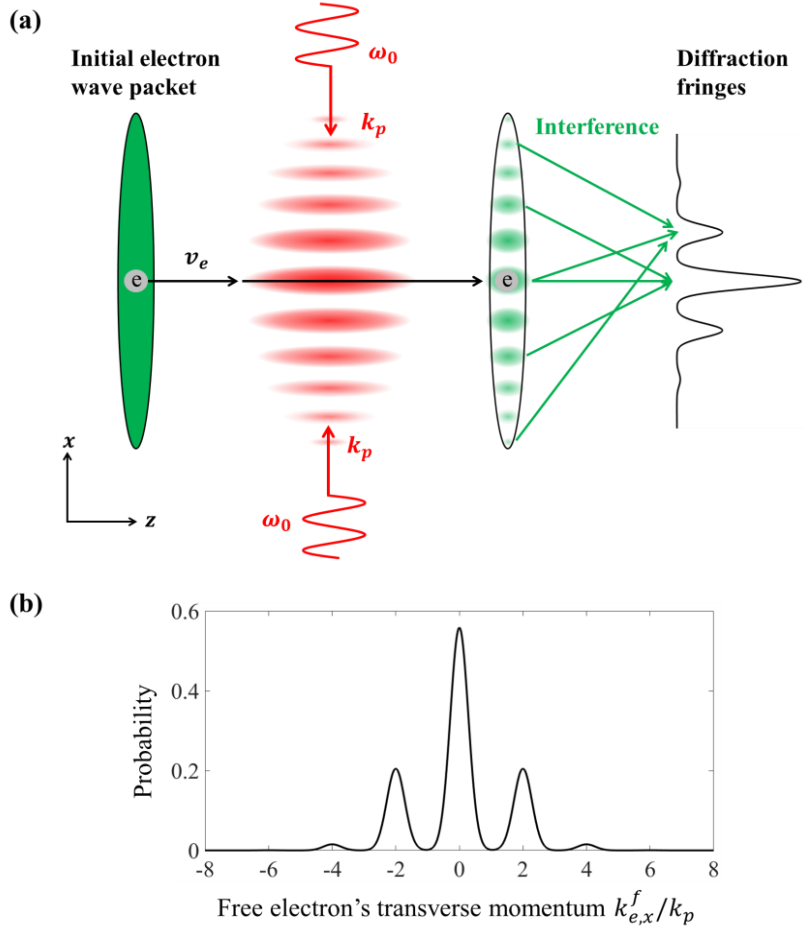

**Fig. 5. The schematic diagram of free electron diffracted by standing light wave and the electron's transverse momentum distribution after interaction.** (a) The initial free electron wave packet is extended along $x$-direction and the different parts of electron wave packet will carry different phases after interaction. The different parts of electron wave packet interference and the diffraction fringes are produced. (b) The calculated probability of transverse momentum of free electrons $k_{e,x}^f$ after interaction. $k_{e,x}^f$ is represented in units of incident photon momentum $k_p$.

From Eq. (17), it can be found that periodic phase is imprinted on the different parts of electron wave packet after interaction, as shown schematically in Fig. 5(a). The different parts of wave packet interfere with each other after free propagation and form the diffraction spots in the detection screen. In Eq. (17), we notice the change of free electrons $k_{e,x}^f$ is $2nk_p$ ($n = 1, 2, 3, \ldots$) along $x$-direction and the corresponding probability is

$$P_n = J_n^2\left(\frac{e^2|E_0|^2 L}{2\gamma m_e \hbar \omega^2 v_e}\right) \quad (18)$$

According to Eq. (18), the probability of electron's transverse momentum is calculated and shown in Fig. 5(b), which is consistent with the theory without quantizing the light field [33] and experiment result [29] of KD effect. Thus, although the phase term $-g_p(\hat{a}\hat{a}^\dagger + \hat{a}^\dagger\hat{a})$ in $\hat{S}$ does not affect the photon number state and the electron energy, it is related to the KD effect and plays an important role in understanding electron diffraction by photons.

Therefore, we explain the KD effect using rigorous QED treatment, verifying it is a phenomenon that free electron is diffracted by emitting/absorbing a photon and absorbing/emitting another photon at the same time. Having understood this mechanism, it is expected to extend the concept of KD effect and manipulate the electron diffraction pattern by introducing photonic structures and using their various optical field characteristics. Thus, the results in this part help further exploring abundant novel phenomenon related to nonlinear Compton scattering.

## 5. Theoretical explanation of the nonlinear Compton scattering

The beforementioned KD effect led to the modulation of transvers momentum of free electron, while the longitudinal momentum remained unchanged. Recently, the inelastic scattering of free electrons when passing through an intense optical traveling wave in vacuum, which generalized the KD effect, was investigated theoretically and experimentally [31-32]. Their experiment results were explained by solving the time-domain Schrödinger equation numerically [32]. In this part, we analyze the experiment using analytical solution of our theory and further explore the entanglement of two photonic states based on this phenomenon.

As depicted in Fig. 6(a), two laser beams with different frequencies $\omega_1$ and $\omega_2$ ($\omega_2 > \omega_1$) were incident at angles $\theta_1$ and $\theta_2$, respectively, in the experiment [32]. These two laser beams form a travelling wave propagating parallel to $z$-axis with group velocity $v_g = (\omega_2 - \omega_1)c/(\omega_2\cos\theta_2 - \omega_1\cos\theta_1)$, which is ensured to be equal to the electron velocity $v_e$ by adjusting the incident angles.

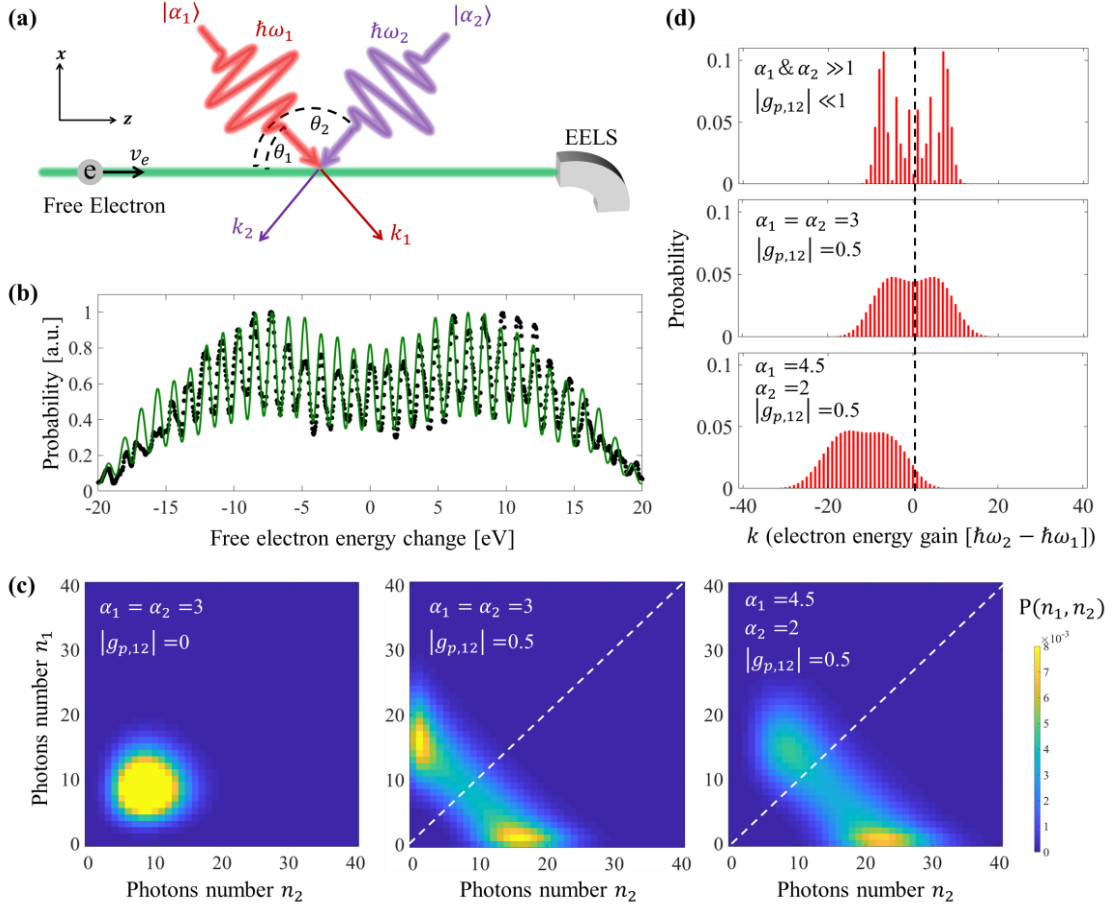

**Fig. 6. The explanation of nonlinear Compton scattering caused by two incident lights with different frequencies. (a)** The free electron beam (green) interact with two coherent photonic states $|\alpha_1\rangle$ (red) and $|\alpha_2\rangle$ (purple), with different frequencies (wave vectors) of $\omega_1$ ($k_1$) and $\omega_2$ ($k_2$), respectively. The after-interaction electrons enter EELS spectrometer and were measured. **(b)** The comparison between measured electron energy spectrum (black dots) from Ref. [32] and fitting curve of analytical expression obtained according to our theory (green curve). The green curve is obtained when the two coherent states are strong while the interaction strength is weak. **(c)** The calculated joint probability distribution $P(n_1, n_2)$ under different strengths of coherence states and $g_{p,12}$. **(d)** The comparison of electron energy spectrums (red bars) under corresponding conditions in (c). The black dot line is the position of zero loss peak (ZLP).

Thus, the Eq. (10) degenerates into $\hat{S} = e^{-2g_{p,12}\hat{b}_1^\dagger \hat{b}_2 \hat{a}_1 \hat{a}_2^\dagger + 2g_{p,12}^* \hat{b}_1 \hat{b}_2^\dagger \hat{a}_1^\dagger \hat{a}_2}$, which has similar form as the phase term discussed in part 4 about explanation of KD effect. It means that they have similar physical process, namely free electron emits/absorbs a photon with energy $\hbar\omega_1$ and absorbs/emits another photon with energy $\hbar\omega_2$ at the same time during interaction. Considering the incident photonic states are two coherent states $|\alpha_1\rangle$ and $|\alpha_2\rangle$, the after-interaction electron-photon states can be written as $\sum_{n_1,n_2=0}^{\infty} \sum_{k=-\infty}^{\infty} c_{n_1 n_2 k} |E_k, n_1, n_2\rangle$, and the coefficient is derived according to the $\hat{S}$ as

$$c_{n_1 n_2 k} = e^{-\frac{|\alpha_1|^2+|\alpha_2|^2}{2}} \frac{\alpha_1^{n_1-k}\alpha_2^{n_2+k}}{(n_1-k)!} e^{-ik\varphi_{gp}} \sum_{m=0}^{\infty} \frac{(-1)^m (\tan 2|g_{p,12}|)^{2m+k}}{(m+k)!\, m!}$$

$$\times (\cos 2|g_{p,12}|)^{2m+n_1-n_2} \frac{(n_1-k+l)!}{(n_2+k-l)!} \sqrt{\frac{n_2!}{n_1!}} \quad (19)$$

where $\varphi_{gp} = \arg(g_{p,12})$. Thus, $|c_{n_1 n_2 k}|^2$ is the probability for free electron to gain energy $k(\hbar\omega_2 - \hbar\omega_1)$ and for the photonic states to have $n_1$ photons with frequency $\omega_1$ and $n_2$ photons with frequency $\omega_2$. For the experimental settings, the coupling strength is weak ($g_{p,12} \ll 1$) and the two coherent states are strong ($\alpha_1 \gg 1$, $\alpha_2 \gg 1$), so that Eq. (19) becomes $c_{n_1 n_2 k} = e^{-ik\varphi_{gp}} J_k(4|g_{p,12}|\sqrt{n_1 n_2}|)$, which has the similar form with single-photon PINEM (more derivation details can be found in Supplementary Material Section S4). Therefore, we can fit the experimental data from Ref. [32] according to our analytical expression, shown as the green curve in Fig. 6(b). The variance of fitness is ~0.012 for our theory, while it is ~0.018 by solving the Schrödinger equation numerically in Ref. [32]. The result means that our theory can well explain the nonlinear Compton scattering and provide analytical expression.

The theory can not only analyze existing experimental phenomena, but also predict new effects for the situation different from the existing experimental settings. After interacting with free electrons, the number states of the two photonic states will be correlated. The joint probability $P(n_1, n_2) = \sum_k |c_{n_1 n_2 k}|^2$ is for indicating the state with $n_1$ photons of frequency $\omega_1$ and $n_2$ photons of frequency $\omega_2$. Assuming the coherent states are weak and the interaction strength is strong, the $P(n_1, n_2)$ is calculated using Eq. (19) and given as color map in Fig. 6(c). When the two coherent states have the same photon number distribution (Poisson distribution, $\alpha_1 = \alpha_2 = 3$) and no interaction with electron ($|g_{p,12}| = 0$), the left figure of Fig. 6(c) illustrates $P(n_1, n_2)$ is derived by the direct multiplication of two Poisson distributions, which indicates the photonic states are noncorrelated. While, $P(n_1, n_2)$ changes significantly for strong $|g_{p,12}| = 0.5$ as shown in the middle figure of Fig. 6(c), indicating that the two photonic states are correlated induced by free electron. When the two incident coherent states have different average number of photons ($\alpha_1 = 4.5$, $\alpha_2 = 2$) and $g_{p,12} = 0.5$, the two entangled peaks are similar to those of the middle figure of Fig. 6(c), but become asymmetric relative to the white dashed line as shown in the right figure of Fig. 6(c).

Although the entanglement between two photonic states induced by free electrons had been reported in recent research [57], it needs two photonic structures and free electrons interacted with these two structures and photonic states one after another. In our scheme, only one photonic structure is need and free electrons interact with two photonic states simultaneously, which is more convenient in structure design for manipulating any two photonic states. What's more, the degree of entanglement is strengthened in our scheme by comparing the two more obvious entanglement peaks with that in Ref. [57]. In short, our theory provides an easier method to induce the entanglement between photons by free electrons.

After interacting with two photonic states, the free electron energy spectrum changes significantly as depicted in Fig. 6(d). When the interaction strength is weak ($|g_{p,12}| \ll 1$), the spectrum has similar distribution as that in previous single-photon PINEM. When increasing interaction strength and two photonic states have the same photon number distribution ($|g_{p,12}| = 0.5$, $\alpha_1 = \alpha_2 = 3$), the envelope of spectrum becomes smoother and the energy spectrum maintains symmetric with respect to zero loss peak (ZLP). When the two incident coherent states have different average number of photons, it is interesting the spectrum has an asymmetric distribution. According to above results, the free electron energy state, for example free electron energy comb, can be modified based on the interaction between free electrons and two photonic states of different frequencies. Moreover, this interaction is hopeful to play an important role in quantum light generation, free electron wave packet manipulation and future quantum informatics.

## IV. Conclusion

In conclusion, the interaction between free electrons and two-photons are investigated systematically by developing the full quantum theory using scattering matrix method. First, defining the second-order quantum coupling constant $g^{(2)}_{qu,ij}$ and $g_{p,ij}$ for representing the two-photon interaction strength, the universal analytical model considering the two-photon process is developed for the first time and the corresponding scattering operator $\hat{S}$ is derived. Second, with the new finding that electron-two-photon interaction is related to not only the longitudinal but also the transverse component of electric field, the $g^{(2)}_{qu,ij}$ could be enhanced by eight orders of magnitude by applying quasi-BIC metasurface instead of previous structures in PINEM, and the full vector optical near field might be derived by the EELS. Third, based on the derived scattering operator $\hat{S}$, the calculated electron-photon joint states indicate that there exists quantum interference between single-photon and two-photon processes, which causes the variation of degree of entanglement between electron and photons, as well as electron energy spectrums, with phase $\Delta\varphi_g$. And the phase of light field can be recovered directly from the EELS of interacted electrons. Finally, based on the developed analytical full quantum theory, the KD effect and nonlinear Compton scattering are also well explained by deducing the probability distribution of electron transverse momentum and electron energy loss, respectively. It is indicated that the single-photon/two-photon PINEM, KD effect and the nonlinear Compton scattering are unified into the derived $\hat{S}$.

This work takes an important step for further study on the free electron-multi-photon interaction in the future, and deepens the understanding of its nonlinear physical mechanism. The feasible scheme to manipulate the wave packet of free electrons by light through two-photon process is expected to be proposed. And the fruitful effects in nonlinear optics based on free electrons may be developed, e.g., realizing spontaneous parametric down-conversion using free electrons as carriers. The concepts presented in this paper are also instructive in the generation of more abundant electron-photon entangled states or quantum photonic states, which are the foundation for quantum sensing, quantum computing and et al.

## ACKNOWLEDGMENT
This work was supported by the National Key Research and Development Program of China (2024YFA1209202).